\documentclass{kapproc} % Computer Modern font calls

\usepackage{procps} 

\usepackage[dvips]{graphicx}

\upperandlowercase
\kluwerbib % will produce this kind of bibliography entry:

\newcommand{\kms}{$\rm km\;s^{-1}$}  
\newcommand{\vsigma}{$V_c$--$\,\sigma_c$}
\newcommand{\sigmac}{$\sigma_c$}
\newcommand{\Vc}{$V_c$}
\newcommand{\Vvir}{$V_{\it vir}$}
\newcommand{\hi}{H~{\small I}} 

\begin{document}

\articletitle[ \vsigma\ relation for LSB galaxies]{The relation 
between circular velocity and central velocity dispersion in low
surface brightness galaxies}

%\articlesubtitle{This is an Article Subtitle}

\author{E.M. Corsini, A. Pizzella, E. Dalla Bonta`, F. Bertola}
\affil{Dipartimento di Astronomia, Universita` di Padova, Italy}
\email{corsini@pd.astro.it}

\author{L. Coccato}
\affil{Kapteyn Astronomical Institute, Groningen, The Netherlands}
%\email{secondauthor@anotheruniv.edu}

\author{M. Sarzi}
\affil{Physics Department, University of Oxford, UK}

\begin{abstract}
We analyzed a sample of high and low surface brightness (HSB and LSB)
disc galaxies and elliptical galaxies to investigate the correlation
between the circular velocity (\Vc ) and the central velocity
dispersion (\sigmac ).
We better defined the previous \vsigma\ correlation for HSB and
elliptical galaxies, especially at the lower end of the
\sigmac\ values.  

Elliptical galaxies with \Vc\ based on dynamical models or directly
derived from the \hi\ rotation curves follow the same relation as the
HSB galaxies in the \vsigma\ plane. On the contrary, the LSB galaxies
follow a different relation, since most of them show either higher
\Vc\ (or lower \sigmac ) with respect to the HSB galaxies.
This argues against the relevance of baryon collapse in the radial
density profile of the dark matter haloes of LSB galaxies.
Moreover, if the \vsigma\ relation is equivalent to one between the
mass of the dark matter halo and that of the supermassive black hole,
these results suggest that the LSB galaxies host a supermassive black
hole with a smaller mass compared to HSB galaxies of equal dark matter
halo. On the other hand, if the fundamental correlation of SMBH mass
is with the halo \Vc , then LSBs should have larger black
hole masses for given bulge \sigmac .
\end{abstract}


\section{Introduction}
\label{sec:introduction}

A possible relation between the central velocity dispersion of the
spheroidal component (\sigmac ) and the galaxy circular velocity (\Vc
) measured in the flat region of the rotation curve (RC) was suggested
by Whitmore et al. (1979). By measuring stellar velocity dispersions
and \hi\ line widths for a sample of 19 spiral galaxies they found a a
significant decrease in \Vc$/$\sigmac\ with increasing bulge-to-disk
ratio. Since \sigmac\ and \Vc\ probe the potential of the spheroid and
dark matter (DM) halo, a mean value \Vc$/$\sigmac\ $\simeq 1.7$
implies these components are dynamically separate with the bulge
substantially cooler than halo.

Gerhard et al. (2001) derived the \vsigma\ relation for the sample of
giant ellipticals studied by Kronawitter et al. (2000). This was
explained as an indication of near dynamical homology of these objects
which were selected to be nearly round and almost non-rotating
elliptical galaxies. These galaxies form a unique dynamical family
which scales with luminosity and effective radius. As a consequence
the maximum \Vc\ of the galaxy is correlated to its
\sigmac . Whether the same is true for more
flattened and fainter ellipticals is still to be investigated.
On the contrary, both shape and amplitude of the RC of a spiral galaxy
depend on the galaxy luminosity and morphological type (e.g., Burstein
\& Rubin 1985). As a consequence for spiral galaxies the \vsigma\
relation is not expected a priori.

Nevertheless, Ferrarese (2002) and Baes (2003) found that elliptical
and spiral galaxies define a common \vsigma\ relation. In particular,
it results that for a given \sigmac\ the value of \Vc\ is independent
of the morphological type. But \sigmac\ and \Vc\ are related to the
mass of the supermassive black hole (hereafter SMBH, see Ferrarese \&
Ford 2005 for a review) and DM halo (e.g., Seljak 2002),
respectively. Therefore Ferrarese (2002) argued that the \vsigma\
relation is suggestive of a correlation between the mass of SMBH and
DM halo.

Previous works concentrated on high surface brightness (HSB)
galaxies. It is interesting to investigate whether the \vsigma\
relation holds also for less dense objects characterized by a less
steep potential well. This is the case of low surface brightness (LSB)
galaxies, which are disc galaxies with a central face-on surface
brightness $\mu_B\geq22.6$ mag arcsec$^{-2}$ (e.g., Impey et
al. 1996). In Pizzella et al. (2005) we studied the behavior of LSB
galaxies in the \vsigma\ relation. Here we present our results.

\section{Sample selection}
\label{sec:sample}

In the past years we started a scientific program aimed at deriving
the detailed kinematics of ionized gas and stars in HSB and LSB
galaxies in order to study their mass distribution and structural
properties (e.g., Pignatelli et al. 2001).

We measured the velocity curves and velocity dispersion profiles along
the major axis for both the ionized-gas and stellar components for a
preliminary sample of $50$ HSB galaxies [$10$ S0--S0/a's in Corsini et
al. (2003); $7$ Sa's in Bertola et al. (1996) and Corsini et
al. (1999); $16$ S0--Sc's in Vega Beltran et al. (2001); $17$
Sb--Scd's in Pizzella et al. (2004a)] and $11$ LSB galaxies (Pizzella
et al. 2004b).
The HSB sample consists of disc galaxies with Hubble type ranging from
S0 to Scd, an inclination $i \geq 30^\circ $ and a distance $D < 80$
Mpc.
The LSB sample consists of disc galaxies with Hubble type ranging from
Sa to Irr, an intermediate inclination ($30^\circ \leq i \leq
70^\circ$), and a distance $D < 65$ Mpc.
In order to complete our sample of disc galaxies we included the 38
Sa-Scd galaxies previously studied by Ferrarese (2002) and the 12
Sb-Sc galaxies by Baes et al. (2003).

Finally, we considered a sample of 24 elliptical galaxies with a flat
RC and for which both \Vc\ and \sigmac\ are available from the
literature. They include 19 objects studied by Kronawitter et
al. (2000) who derived \Vc\ by dynamical modeling and 5 objects
studied by Bertola et al. (1993) for which \Vc\ is directly derived
from the flat portion of their \hi\ RCs. The addition of these
galaxies will allow to test against model-dependent biases in the
\vsigma\ relation.

\section{Measuring \Vc\ and \sigmac }

For all the disc galaxies we obtained the ionized-gas RC. We rejected
galaxies with asymmetric RCs or RCs which were not characterized by an
outer flat portion. The flatness of each RC has been checked by
fitting it with a linear law $V(R)=AR+B$ for
$R\;\geq\;0.35\;R_{25}$. The radial range has been chosen in order to
avoid the bulge-dominated region. The RCs with $|A|\geq2$ $\rm
km\;s^{-1}\;kpc^{-1}$ within 3$\sigma$ have been considered not to be
flat. We derived \Vc\ by averaging the outermost values of the flat
portion of the RC.  We derived \sigmac\ from the stellar kinematics by
extrapolating the velocity dispersion radial profile to the
centre. This has been done by fitting the innermost data points with
an empirical function (either an exponential, or a Gaussian or a
constant). We did not apply any aperture correction to
\sigmac\ as discussed by Pizzella et al. (2004a).

For the elliptical galaxies we obtained \Vc\ either from the dynamical
models by Gerhard et al. (2001) or from the flat portion of the \hi\
RC. For these galaxies we relaxed the flatness criterion in favor of
their large radial extension which is about 10 times larger than that
of optical RCs. Aperture measurements of the stellar velocity
dispersion corrected to $r_e/8$ were adopted to estimate \sigmac .

\section{The \vsigma\ relation for HSB and elliptical galaxies}

In summary, we have $40$ disc galaxies [$15$ from our preliminary
sample, $16$ from Ferrarese (2002), and $9$ from Baes et al. (2003)
with flat RCs extending to $\sim\;0.8\;R_{25}$] and $24$ elliptical
galaxies [$19$ from Kronawitter et al. (2000) with flat RCs extending
to about $\sim\;0.5\;R_{25}$, $5$ from Bertola et al. (1993) with flat
\hi\ RCs extending to $\sim\;3\;R_{25}$].

A power law has been usually adopted to describe the correlation
between \Vc\ and \sigmac\ for galaxies with \sigmac$>80$
\kms . We find
\begin{equation}
\log V_c  = (0.74\pm 0.07)\, \log \sigma_c + (0.80 \pm 0.15)  
\label{eq:HSBfit_power}
\end{equation}
with \Vc\ and \sigmac\ expressed in \kms. The resulting power law is
plotted in Fig. 1 (dash-dotted line). However, according to a $\chi^2$
analysis the data are also consistent with a linear law out to
velocity dispersions as low as \sigmac$\;\approx50$ \kms\
\begin{equation}
V_c = (1.32\pm 0.09)\,\sigma_c + (46\pm14).
\label{eq:HSBfit_linear}
\end{equation}
The resulting straight line is plotted in Fig. 1 (continuous line).

The data points corresponding to the 5 elliptical galaxies with \Vc\
based on \hi\ data follow the same relation as the remaining disc and
elliptical galaxies.
They are mostly located on the upper end of the
\vsigma\ relation derived for disc galaxies, in agreement 
with the findings of Bertola et al. (1993). They studied these
elliptical galaxies and showed that their DM content and distribution
are similar to those of spiral galaxies.

\begin{figure}[t]
\begin{center}
\includegraphics[angle=0,width=9.5cm]{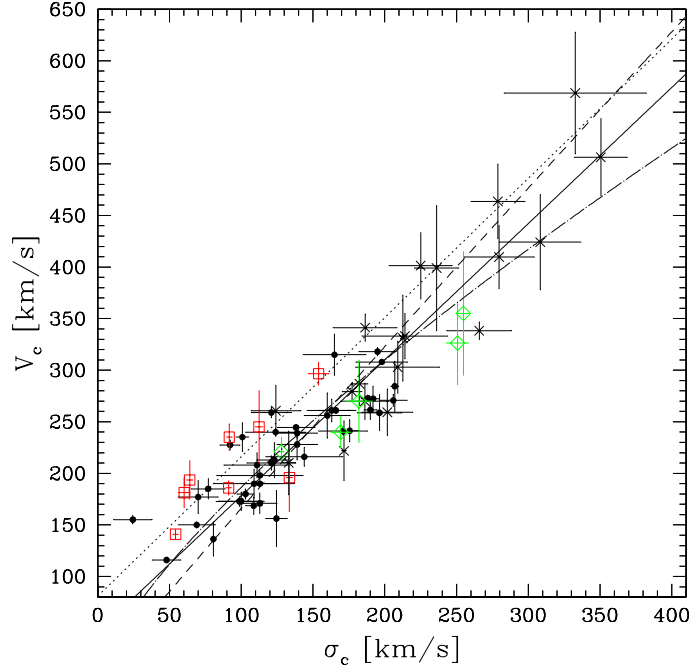}
\end{center}
\caption{The correlation between \Vc\ and \sigmac\. The data points
  corresponding to HSB galaxies ({\em filled circles}), LSB galaxies
  ({\em squares}), and elliptical galaxies with \Vc\ obtained from
  \hi\ data ({\em diamonds}) or dynamical models ({\em crosses}) are
  shown. The {\em dash-dotted} and {\em continuous line} represent the
  power-law (Eq. 1) and linear fit (Eq. 2) to HSB and elliptical
  galaxies. The {\em dotted line} represents the linear-law fit
  (Eq. 3) to LSB galaxies. For a comparison, the {\em dashed line}
  corresponds to the power-law fit to spiral galaxies with
  $\sigma_c>80$ km s$^{-1}$ by Baes et al. (2003).}
\end{figure}

\section{The \vsigma\ relation for LSB galaxies}

The LSB and HSB galaxies do not follow the same \vsigma\ relation. In
fact, most of the LSB galaxies are characterized by a higher \Vc\ for
a given \sigmac\ (or a lower \sigmac\ for a given \Vc ) with respect
to HSB galaxies. By applying to the LSB data points the same
regression analysis which has been adopted for the HSB and elliptical
galaxies of the final sample, we find
\begin{equation}
V_c  = (1.35\pm 0.19)\,\sigma_c + (81\pm23)  
\label{eq:LSBfit_linear}
\end{equation}
with \Vc\ and \sigmac\ expressed in \kms . The straight line
corresponding to this fit, which is different from the one obtained
for HSB and elliptical galaxies and happens to be parallel to it, is
plotted in Fig. 1 (dotted line). To address the significance of this
result, which is based only on 8 data points, we compared the
distribution of the normalized scatter of the LSB galaxies to that of
the HSB and elliptical galaxies.  They are different at a high
confidence level ($>99\%$) by a Kolmogorov-Smirnov test. The fact that
these objects fall in a different region of the \vsigma\ plane
confirms that LSB and HSB galaxies constitute two different classes of
galaxies.

\section{Conclusions}

Both demographics of SMBHs and study of DM distribution in galactic
nuclei benefit from the \vsigma\ relation.  The recent finding that
the mass of SMBHs correlates with different properties of the host
spheroid supports the idea that formation and accretion of SMBHs are
closely linked to the formation and evolution of their host
galaxy.
A task to be pursued is to obtain a firm description of all these
relationships spanning a wide range of SMBH masses and address if they
hold for all Hubble types. In fact, the current demography of SMBHs
suffers of important biases, related to the limited sampling over the
different basic properties of their host galaxies.
The finding that the \vsigma\ relation holds for small values of
\sigmac\ points to the idea that SMBHs with masses 
M$_\bullet<10^6$ M$_\odot$ may also exist and follow the
M$_\bullet$-$\sigma$ relation.

Moreover, it has been suggested that the \vsigma\ relation is
equivalent to one between the masses of SMBH and DM halo (Ferrarese
2002; Baes 2003) because \sigmac\ and \Vc\ are related to the masses
of the central SMBH and DM halo, respectively. Yet, this claim is to
be considered with caution, as the demography of SMBHs is still
limited, in particular as far as spiral galaxies are concerned.
Furthermore, the calculation of the virial mass of the DM halo from
the measured \Vc\ depends on the assumptions made for the DM density
profile and the resulting rotation curve (e.g., see the prescriptions
by Seljack et al. 2002). A better estimate of the virial velocity of
the DM halo \Vvir\ can be obtained by constraining the
baryonic-to-dark matter fraction with detailed dynamical modeling of
the sample galaxies. The resulting \Vvir$-$\sigmac\ relation is
expected to have a smaller scatter than the \vsigma\ relation.
If the M$_\bullet$-$\sigma$ relation is to hold, the
deviation of LSB galaxies {\it with bulge\/} from the \vsigma\ of HSB
and elliptical galaxies suggests that for a given DM halo mass the LSB
galaxies would host a SMBH with a smaller mass compared to HSB
galaxies.
On the other hand, if the fundamental correlation of SMBH mass is with
the halo \Vc , then LSBs should have larger black-hole
masses for given bulge \sigmac .  This should be accounted
for in the theoretical and numerical investigations of the processes
leading to the formation of LSB galaxies.

The collapse of baryonic matter can induce a further concentration in
the DM distribution, and a deepening of the overall gravitational well
in the central regions. If this is the case, the finding that at a
given DM mass (as traced by \Vc ) the central \sigmac\ of LSB galaxies
is smaller than in their HSB counterparts, would argue against the
relevance of baryon collapse in the radial density profile of DM in
LSB galaxies.

\end{document}